\documentclass[aps,pra,twocolumn]{revtex4} 

\usepackage{amsmath}    % need for subequations
\usepackage{graphicx}   % for figures
\usepackage[usenames,dvipsnames]{xcolor}

\begin{document}
 
\newcommand{\ket}[1]{|#1\rangle}
\newcommand{\bra}[1]{\langle#1|}
\newcommand{\overl}[2]{\langle#1|#2\rangle}
\newcommand{\ph}{|\phi\rangle}
\newcommand{\ps}{|\psi\rangle}

\newcommand{\0}{|0\rangle}
\newcommand{\1}{|1\rangle}
\newcommand{\+}{\frac{\0+\1}{\sqrt{2}}}
\newcommand{\m}{\frac{\0-\1}{\sqrt{2}}}
\newcommand{\Cbas}{\{\0,\1 \}}
\newcommand{\Hbas}{\{\+,\m \}}

\title{The SWAP test and the Hong-Ou-Mandel effect are equivalent}
\author{Juan Carlos Garcia-Escartin}

 \affiliation{Universidad de Valladolid, Dpto. Teor\'ia de la Se\~{n}al e Ing. Telem\'atica, Paseo Bel\'en n$^o$ 15, 47011 Valladolid, Spain}
\email{juagar@tel.uva.es}   %optional
\author{Pedro Chamorro-Posada}
\affiliation{Universidad de Valladolid, Dpto. Teor\'ia de la Se\~{n}al e Ing. Telem\'atica, Paseo Bel\'en n$^o$ 15, 47011 Valladolid, Spain}
\date{\today}

\begin{abstract}
We show that the Hong-Ou-Mandel effect from quantum optics is equivalent to the SWAP test, a quantum information primitive which compares two arbitrary states. We first derive a destructive SWAP test that doesn't need the ancillary qubit that appears in the usual quantum circuit. Then, we study the Hong-Ou-Mandel effect for two photons meeting at a beam splitter and prove it is, in fact, an optical implementation of the destructive SWAP test. This result offers both an interesting simple realization of a powerful quantum information primitive and an alternative way to understand and analyse the Hong-Ou-Mandel effect.
\end{abstract}
\maketitle

\section{Introduction}
\label{intro}
Quantum information has provided a new way to think about quantum mechanics. Its formalism draws heavily from quantum optics and many interesting results come from the interplay between both disciplines. Bell inequalities and Bell tests can be more clearly understood in a computational framework \cite{BCM10}. Simple quantum information protocols, such as quantum cryptography are naturally realized with optical systems \cite{BB84,GRT02}. Many quantum algorithms are also directly inspired by physical phenomena. For instance, Grover's algorithm for quantum search is based on Schr\"odinger's Equation \cite{Gro01}.

In this paper, we show how quantum information has ``rediscovered'' the Hong-Ou-Mandel effect of quantum optics under the name of SWAP test. We show there is a deep connection between these two concepts. On the way, we propose a new SWAP test circuit that doesn't need any ancillary inputs and suggest practical realizations of this test using photons, a beam splitter and two detectors. 

The paper has five main sections. In Section \ref{SWAP}, we describe the SWAP test and its uses in state comparison. In Section \ref{HOM}, we review the Hong-Ou-Mandel effect for two photons and give a formulation that highlights the role of  the information the photons carry. In Section \ref{dSWAP}, we derive a destructive SWAP test circuit with no ancillas. Section \ref{opSWAP} shows the Hong-Ou-Mandel effect corresponds to a destructive, simplified optical SWAP test circuit. Finally, in Section \ref{apps}, we outline the possible applications of these results and propose experimental systems that put these connections into practical use. 

\section{The SWAP test}
\label{SWAP}
\begin{figure}[ht!]
\centering
\includegraphics[scale=1]{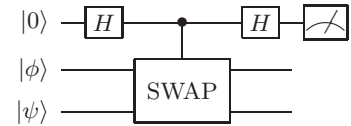}
\caption{Quantum circuit implementing the SWAP test.\label{SWAPcirc}} 
\end{figure}

When working with quantum information, it often appears the question of whether two states $\ph$ and $\ps$ are equal or not. The SWAP test is a procedure from which we can determine with certainty that two states are different. Equality can be inferred with high probability if we have multiple copies of the states. The quantum circuit used in the test, introduced in the context of quantum fingerprinting \cite{BCW01}, is shown in Figure \ref{SWAPcirc}.
The inputs are two states $\ps$ and $\ph$ of equal dimension and an ancillary qubit in the $\0$ state. There are three gates, two Hadamard gates, $H$, and a controlled SWAP gate, CSWAP. The Hadamard gates convert the $\0$ state into a superposition $\+$ and $\1$ into $\m$. The controlled SWAP operation interchanges the states $\ph$ and $\ps$ if the ancillary qubit is in state $\1$. When the ancillary qubit is $\0$, the other states keep their order. The evolution through this circuit is:
\begin{equation}
\0\ph\ps \stackrel{\footnotesize{H}}{\rightarrow}\frac{\0+\1}{\sqrt{2}}\ph\ps\stackrel{{\tiny C\!SW\!AP}}{\longrightarrow}\frac{\0\ph\ps+\1\ps\ph}{\sqrt{2}}\nonumber
\end{equation}
\begin{equation}
\stackrel{\footnotesize{H}}{\rightarrow}\frac{\0\left[\ph\ps+\ps\ph\right]+\1\left[\ph\ps-\ps\ph\right]}{2}.\label{SWAPout}
\end{equation}
At the end of the circuit, the state of the ancillary qubit is measured. We call outcome 0 the case where the $\0$ state is found and outcome 1 when $\1$ is measured.
If the states are equal, $\ph=\ps$, the outcome is 0 with probability 1. Swapping the positions has no effect and there is no entanglement with the ancillary qubit. For different states both outcomes are possible. Outcome 1 can only happen for different states. In that case, we say the states ``fail'' the test. If two states fail the test, we know with certainty they are different. If the states ``pass'' the test (outcome 0), they are not necessarily equal. From Equation (\ref{SWAPout}), we can find the probability for passing the test is:
\begin{equation}
P=\frac{1}{4}(\bra{\phi}\bra{\psi}+\bra{\psi}\bra{\phi})(\ph\ps+\ps\ph)=\frac{1+|\overl{\psi}{\phi}|^2}{2}.
\end{equation}  
The probability of failure is the complementary $\frac{1-|\overl{\psi}{\phi}|^2}{2}$. The test is only valid as a comparison of independent input states. If the inputs are entangled, the state must be taken as a whole and it makes no sense to speak of a comparison.

The probability of passing the test depends on the overlap $|\overl{\psi}{\phi}|^2$ of the input states. The overlap gives a good estimate of how close two states are. For two orthogonal states, $|\overl{\psi}{\phi}|^2=0$ and $P=\frac{1}{2}$. For non-orthogonal states, the closer they are, the greater the probability of passing the test. If we have $n$ copies of the two input states, we can repeat the test. The probability of passing the $n$ rounds is 
\begin{equation}
\left(\frac{1+|\overl{\psi}{\phi}|^2}{2}\right)^n.
\end{equation}
If the state passes the test multiple times, we can infer with high probability they are equal or, at least, have a small overlap. We can estimate the number of tests we need to tell apart two states which are arbitrarily close so that $|\overl{\psi}{\phi}|^2= 1 - \epsilon$, with $\epsilon \ll 1$. With this overlap, $P=\frac{2-\epsilon}{2}$ and the probability of passing $n$ tests is $(1-\frac{\epsilon}{2})^n\approx 1-\frac{n\epsilon}{2}$.

One important detail of the SWAP test is the output state after measuring the ancillary qubit. For outcome 0, we have an entangled state $\0\frac{\ph\ps+\ps\ph}{\sqrt{2}}$ and for outcome 1, $\1\frac{\ph\ps-\ps\ph}{\sqrt{2}}$. In both cases, it is impossible to completely separate the input states again. If it were possible, the SWAP test could be repeated as many times as desired. This would allow to distinguish arbitrarily close states. It is easy to see why this must be wrong. If the states could be recycled, we could choose a set of states $\left\{ \ket{\psi_1}, \ket{\psi_2},\ldots, \ket{\psi_N} \right\}$ as large as we want and, for an unknown state $\ket{\psi_i}$, we could try each of them until we find an outcome 1. After a defined number of tries, we would deduce with high probability index $i$. This method allows to send an arbitrarily large amount of information encoded in a state of a finite dimension. This clearly violates the Holevo bound, which gives a limit of $\log_2 d$ bits for a $d$-dimensional system \cite{Hol73,Hol98}.

Due to this confusion of states at the output, the protocols that use the SWAP test do no further work on them. The output can be measured without any effect on the protocol. This motivates our search for a simpler test with no ancillary qubit and where the output is measured destroying any superposition. In section \ref{dSWAP}, we describe an ancilla-free test using standard quantum gates, but, first, we show a simple optical system which already gives a destructive quantum state comparison.

\section{The Hong-Ou-Mandel effect}
\label{HOM}
The Hong-Ou-Mandel (HOM) effect of quantum optics offers a straightforward way to compare the state of two photons. The phenomenon was originally proposed as a way to find nanosecond timeshifts between two photons \cite{HOM87}, but, in its full generality, it can help to detect any other difference, like frequency shifts or other changes in the wavefunction. 

We can describe the phenomenon by looking at the behaviour of photons when they cross a beam splitter. We imagine a photon in state $\ket{s}$ which can take two paths, up and down. We use the notation $\ket{s_p^n}$ to denote a photon number state $\ket{n}$ in mode $s_p$. Mode $s_p$ describes a photon with a certain state $\ket{s}$ which can include polarization or frequency, while subindex $p$ is reserved to specify the path (spatial mode), which can be up  $\ket{s_U}$ or down $\ket{s_D}$. The vacuum state (zero photon number) is represented as $\ket{0_U}$ or $\ket{0_D}$. All the modes have the same vacuum state (all empty modes are the same).

For a 50\% beam splitter we have the evolution:
\begin{equation}
\ket{s_U^1}\ket{0_D} \longrightarrow \frac{\ket{s_U^1}\ket{0_D}+\ket{0_U}\ket{s_D^1}}{\sqrt{2}}
\end{equation}
and
\begin{equation}
\ket{0_U}\ket{s_D^1} \longrightarrow \frac{\ket{s_U^1}\ket{0_D}-\ket{0_U}\ket{s_D^1}}{\sqrt{2}}.
\end{equation}
For single photons, this is the equivalent of an $H$ gate where we replace logic states $\0$ and $\1$ by $\ket{s_U^1}\ket{0_D}$ and $\ket{0_U}\ket{s_D^1}$ respectively. If we place two detectors $D_1$ and $D_2$, one up and one down, each can ``click'' (find the photon) with a probability $\frac{1}{2}$. 

If we have two photons in orthogonal modes $\ket{s^1}$ and $\ket{t^1}$, with $\overl{s^1}{t^1}=0$, one up and one down, they evolve independently. The final click statistics in the detectors can be deduced from those of the individual photons. When $D_1$ and $D_2$ click at the same time, or, in practice, in the same short time window, we say there is a \emph{coincidence}. An interesting phenomenon appears when the input photon states have an overlap $\overl{s^1}{t^1}\neq0$. Photons in the same state bunch together at the output. The simplest case occurs for two indistinguishable input photon states. 

We can describe the general evolution inside a beam splitter or any other linear optics element from its scattering matrix. We use photon creation operators $\hat{a}^\dag_{s,p}$ such that
\begin{equation}
\label{creation}
\hat{a}^\dag_{s,p}\ket{s_p^n}=\sqrt{n+1}\ket{s_p^{n+1}}.
\end{equation}
A state $\ket{s_p^n}$ can be written as \cite{Lou00}
\begin{equation}
\ket{s_p^n}=\frac{(\hat{a}^\dag_{s,p})^n}{\sqrt{n!}}\ket{0_p}.
\end{equation}

The creation operators of independent modes (orthogonal photon states) commute. In the Heisenberg picture, we can study the evolution of a quantum system from the evolution of an operator acting on the same initial state. If the evolution is defined by a unitary operator $U$ and we have an input photon in state $\ket{s_p^1}=\hat{a}_{s,p}^\dag\ket{0_p}$, the output after the beam splitter can be written as $(U \hat{a}_{s,p}^\dag U^\dag)\ket{0_p}$. We concentrate on the evolution of the operator. A 50\% beam splitter has a scattering matrix
\begin{equation}
S=\left( \begin{array}{cc}
\frac{1}{\sqrt{2}} & \frac{1}{\sqrt{2}} \\
\frac{1}{\sqrt{2}} & -\frac{1}{\sqrt{2}} \end{array} \right)
\end{equation}
and it can be shown that the creation operators evolve as \cite{SGL04}:
\begin{eqnarray}
\label{BSU}
U\hat{a}_{s,U}^\dag U^\dag &\longrightarrow& \frac{1}{\sqrt{2}} \hat{a}_{s,U}^\dag+\frac{1}{\sqrt{2}} \hat{a}_{s,D}^\dag\\
U\hat{a}_{s,D}^\dag U^\dag &\longrightarrow& \frac{1}{\sqrt{2}} \hat{a}_{s,U}^\dag-\frac{1}{\sqrt{2}} \hat{a}_{s,D}^\dag.
\label{BSD}
\end{eqnarray}
For two equal photons giving an input state 
\begin{equation}
\ket{s_U^1}\ket{s_D^1}=\hat{a}_{s,U}^\dag\hat{a}_{s,D}^\dag\ket{0_U}\ket{0_D},
\end{equation}
we have at the output
\begin{multline}
U\hat{a}_{s,U}^\dag\hat{a}_{s,D}^\dag U^\dag\ket{0_U}\ket{0_D}=(U\hat{a}_{s,U}^\dag U^\dag)(U\hat{a}_{s,D}^\dag U^\dag)\ket{0_U}\ket{0_D}\\
=\frac{1}{2}((\hat{a}_{s,U}^\dag)^2-\hat{a}_{s,U}^\dag\hat{a}_{s,D}^\dag+\hat{a}_{s,D}^\dag\hat{a}_{s,U}^\dag-(\hat{a}_{s,D}^\dag)^2)\ket{0_U}\ket{0_D}.
\end{multline}
For modes $U$ and $D$ the creation operators commute and the output state is
\begin{equation}
\frac{(\hat{a}_{s,U}^\dag)^2-(\hat{a}_{s,D}^\dag)^2}{2}\ket{0_U}\ket{0_D},
\end{equation}
which, from (\ref{creation}), is
\begin{equation}
\frac{\ket{s_U^2}\ket{0_D}-\ket{0_U}\ket{s_D^2}}{\sqrt{2}}.
\end{equation}
Due to interference, both photons leave the beam splitter through the same port. Both detectors have an equal probability of clicking, but the number of coincidences becomes zero. 

For photons with continuous wavepacket amplitude functions $\xi_1(t)$ and $\xi_2(t)$ at the input of a 50\% beamsplitter, the probability of finding a coincidence is known to be \cite{Lou00}:
\begin{equation}
\label{contHOM}
\frac{1-| \int  \xi_1(t)^* \xi_2(t)\,dt|^2}{2}.
\end{equation}
The term $| \int  \xi_1(t)^* \xi_2(t)\,dt|$ is the overlap of the two photon states. The photons are found in the same output port with the same probability
\begin{equation}
\frac{1+| \int  \xi_1(t)^* \xi_2(t)\,dt|^2}{4}
\end{equation}
for each detector. The photons have clearly the same behaviour as the input states in the SWAP test. In the following section, we derive the same results for discrete systems which correspond naturally to the qubit or qudit case. We study the system from the point of view of discrete photon creation operators. Later, we discuss its equivalence to the general expression with the wavepacket amplitude functions. 

\subsection{Discrete systems. HOM for d-dimensional systems}
\label{disc}
We want to consider now photon states $\ph=\sum_{i=0}^{d-1}\alpha_i\ket{i}$ and $\ps=\sum_{j=0}^{d-1}\beta_j\ket{j}$. States $\ket{i}$ from $\{ \ket{0},\ket{1}, \ldots,\ket{d-1}\}$ are orthogonal and can correspond to photons with different frequencies, orbital angular momentum values \cite{MTT01} or with wavefunctions in different time windows like in time-bin encoding \cite{SZG05}.

We have now creation operators $\hat{a}_{i,p}^\dag$, as we still allow each of these photon states to be in the upper and lower ports. The evolution through a 50\% beam splitter is
\begingroup
\everymath{\small}
\small
\begin{multline}
\ket{\phi_U}\ket{\psi_D}=\sum_{i}^{d-1}\sum_{j}^{d-1}\alpha_i\beta_j\hat{a}_{i,U}^\dag\hat{a}_{j,D}^\dag\ket{0_U}\ket{0_D}\\
\rightarrow U\left(\sum_{i}\sum_{j}\alpha_i\beta_j\hat{a}_{i,U}^\dag\hat{a}_{j,D}^\dag\right)U^\dag\ket{0_U}\ket{0_D}\\
=\sum_{i}\sum_{j}\alpha_i\beta_j(U\hat{a}_{i,U}^\dag U^\dag)(U\hat{a}_{j,D}^\dag U^\dag)\ket{0_U}\ket{0_D}\\
\!=\!\sum_{i}\!\sum_{j}\!\frac{\alpha_i\beta_j}{2}\!(\hat{a}_{i,U}^\dag\hat{a}_{j,U}^\dag\!-\hat{a}_{i,U}^\dag\hat{a}_{j,D}^\dag\!+\hat{a}_{i,D}^\dag\hat{a}_{j,U}^\dag\!-\hat{a}_{i,D}^\dag\hat{a}_{j,D}^\dag)\ket{0_U}\ket{0_D}.
\end{multline}
\endgroup

There are two parts with different behaviour
\begingroup
\everymath{\small}
\small
\begin{multline}
\sum_{i}\alpha_i\beta_i\frac{\ket{i_U^2}\ket{0_D}-\ket{0_U}\ket{i_D^2}}{\sqrt{2}}+\\
\sum_{i}\sum_{j\neq i}\frac{\alpha_i\beta_j}{2}[\ket{i_U^1}\ket{j_U^1}\ket{0_D}-\ket{i_U^1}\ket{j_D^1}+\ket{j_U^1}\ket{i_D^1}-\ket{0_U}\ket{i_D^1}\ket{j_D^1}].
\label{pass}
\end{multline}
\endgroup

We can consider the setting as a test. The photons pass the test if only one detector fires. A coincidence is detected as a failure. The probabilities of each event are related to the overlap of the two input states, with $\overl{\psi}{\phi}=\sum_{i}\alpha_i\beta_i^*$ and
\begin{equation}
\label{overlapeq}
|\overl{\psi}{\phi}|^2=\overl{\psi}{\phi}\overl{\psi}{\phi}^*=\sum_{i}\sum_{j}\alpha_i\alpha_j^*\beta_i^*\beta_j.
\end{equation}

The part of the superposition in (\ref{pass}) which corresponds to a coincidence is:
\begin{equation}
\sum_{i}\sum_{j\neq i}\frac{\alpha_i\beta_j}{2}[-\ket{i_U^1}\ket{j_D^1}+\ket{j_U^1}\ket{i_D^1}].
\end{equation}
The terms can be rearranged taking into account the interference between indistinguishable photon states to give
\begin{equation}
\sum_{i}\sum_{j}\frac{1}{2}(\alpha_i\beta_j-\alpha_j\beta_i)\ket{i_U^1}\ket{j_D^1}.
\end{equation}

The probability of finding a coincidence and failing the test is
\begin{multline}
\sum_{i}\sum_{j}\frac{1}{4}(\alpha_i\beta_j-\alpha_j\beta_i)(\alpha_i\beta_j-\alpha_j\beta_i)^*=\\
\sum_{i}\sum_{j}\frac{1}{4}(|\alpha_i|^2|\beta_j|^2+|\alpha_j|^2|\beta_i|^2-\alpha_i\alpha_j^*\beta_i^*\beta_j- \alpha_i^*\alpha_j\beta_i\beta_j^*).
\end{multline}
We can group the terms and see the failure probability is
\begin{equation}
\frac{\sum_{i}\sum_{j}|\alpha_i|^2|\beta_j|^2-\sum_{i}\sum_{j}\alpha_i\alpha_j^*\beta_i^*\beta_j}{2}=
\frac{1-|\overl{\phi}{\psi}|^2}{2},
\end{equation}
where we use that $\sum_{i}|\alpha_i|^2=1$, $\sum_{i}|\beta_i|^2=1$ and Equation (\ref{overlapeq}).

The probability of passing the test is, as it should, $\frac{1+|\overl{\phi}{\psi}|^2}{2}$ for any pair of input states $\ph$ and $\ps$. This shows the Hong-Ou-Mandel circuit performs a SWAP test. The formal equivalence permits an optical implementation in many applications where the SWAP test is used (see section \ref{apps}). 

\section{Destructive SWAP tests}
\label{dSWAP}
The Hong-Ou-Mandel effect proves no ancillary photon is needed to perform a SWAP test. In this Section, we present a destructive SWAP test with no ancillas.

We can derive the new circuit from an implementation of the SWAP gate which only uses CNOT and Toffoli gates. Both are based on the binary exclusive or function (XOR). The XOR of two binary values is only true if one, and only one, of them is true. Table \ref{XORtable} shows the XOR truth table.

\begin{table}[h!]
\vspace{2.5ex}
\begin{center}
\begin{tabular}{| cc | c |}	
\hline
\multicolumn{3}{|c|}{\bf{XOR}}\\
\hline
\phantom{a}x\phantom{a}&\phantom{a}y\phantom{a}&\phantom{a}$x\oplus y$\phantom{a}\\
\hline
\colorbox{gray}{\emph{0}}&\bf{\emph{0}}&\bf{\emph{0}}\\
\colorbox{gray}{0}&\bf{1}&\bf{1}\\
\hline
1&0&1\\
\emph{1}&\emph{1}&\emph{0}\\
\hline
\end{tabular}
\caption{Truth table for the XOR logical operation.}
\label{XORtable}
\end{center}
\end{table}

In particular, we use that $x\oplus x=0$ and $x \oplus 0=x$ for any input. The XOR function can be seen as both a modulo 2 addition and a modulo 2 complement. In higher dimensions these operations correspond to separate functions. 

In our gates, we define target and control qubits represented by the XOR symbol ($\oplus$) and a black dot respectively. The CNOT gate is a controlled NOT operation that flips the target value if the control is in state $\1$. We have
\begin{equation}
\mbox{CNOT}\ket{x}\ket{y}=\ket{x}\ket{x\oplus y}.
\end{equation} 
The Toffoli gate is a controlled-controlled-NOT. The target is only flipped if both control qubits are $\1$, with evolution
\begin{equation}
\mbox{CCNOT}\ket{x}\ket{y}\ket{z}=\ket{x}\ket{y}\ket{(x\cdot y)\oplus z},
\end{equation} 
where $x\cdot y$ is the binary AND of $x$ and $y$. From the properties of the XOR function, we can see both gates are their own inverses. They cancel if applied twice in a row. 
 
\subsection{Comparison of one qubit states}
\begin{figure}[ht!]
\centering
\includegraphics[scale=1]{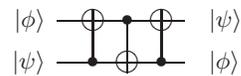}
\caption{XOR swapping circuit.\label{XORSWAP}} 
\end{figure}

We will start with a CSWAP circuit inspired by classical XOR swapping. When we have two registers, we can switch their contents without any additional memory bits with the circuit of Figure \ref{XORSWAP}.

The step-by-step evolution is 
\begin{multline}
\ket{x}\ket{y}\stackrel{\mbox{\scriptsize{CNOT}}(2,1)}{\longrightarrow}\ket{x\oplus y}\ket{y}\stackrel{\mbox{\scriptsize{CNOT}}(1,2)}{\longrightarrow}\ket{x\oplus y}\ket{y\oplus x \oplus y}=\\
\ket{x\oplus y}\ket{x }\stackrel{\mbox{\scriptsize{CNOT}}(2,1)}{\longrightarrow}\ket{x\oplus y \oplus x}\ket{x}=\ket{y}\ket{x}. 
\end{multline}
We call $\mbox{CNOT}(i,j)$ the CNOT gate with control qubit $i$ and target qubit $j$. We have described the classical setting, but the results can also be generalized to arbitrary quantum superpositions.

\begin{figure}[ht!]
\centering
\includegraphics[scale=1]{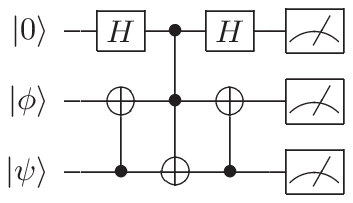}
\caption{SWAP test for qubit states with a CSWAP gate.\label{CSWAP}} 
\end{figure}

For the CSWAP gate we introduce an additional control in the middle gate (see the SWAP test circuit of Figure \ref{CSWAP}). If the ancillary control qubit is $\0$, the middle gate has no effect and the side CNOTs cancel. If the ancillary qubit is $\1$, we recover the SWAP operation.

\begin{figure}[ht!]
\centering
\includegraphics[scale=1]{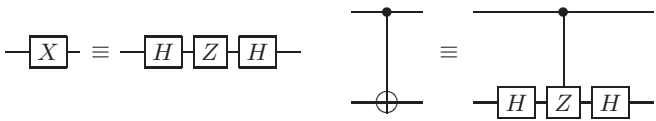}
\caption{Equivalent circuits for the X and Z gates and their controlled versions.\label{circuits}} 
\end{figure}

We are going to find equivalent circuits to show the ancillary qubit can be replaced by measurement on the tested states. We are going to use the equivalences of Figure \ref{circuits}.

A NOT gate, also called $X$, is equivalent to a controlled sign shift $Z$ such that $Z\ket{x}=(-1)^x\ket{x}$ surrounded by two Hadarmard gates. The $Z$ gate introduces a minus sign in the $\1$ terms of the qubit superposition. The first $H$ gate takes the state into the $\Hbas$ basis, where the $Z$ gate acts as a NOT operation (takes one of the basis states to the other). When we go back to the computational, $\Cbas$, basis, the states are flipped. 

We can immediately see that, if we replace the $Z$ gate by a controlled $Z$, CZ gate such that $\mbox{CZ}\ket{x}\ket{y}=(-1)^{x\cdot y}\ket{x}\ket{y}$, the resulting circuit acts as a CNOT gate. If the control qubit is $\1$, we have the operation sequence $HZH=X$. If it is $\0$, we have two $H$ gates which cancel. We can similarly define a controlled-controlled-$Z$ gate, CCZ, with $\mbox{CCZ}\ket{x}\ket{y}\ket{z}=(-1)^{x\cdot y\cdot z}\ket{x}\ket{y}\ket{z}$. With these equivalences, we can proceed to simplify the SWAP test circuit.

\begin{figure}[ht!]
\centering
\includegraphics[scale=1]{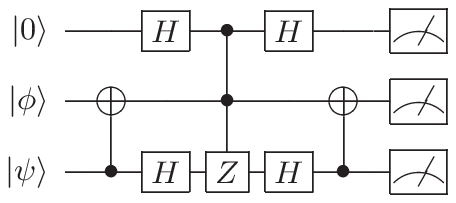}
\caption{SWAP test circuit with a CCZ gate.\label{swap3}} 
\end{figure}

Figure \ref{swap3} shows our starting circuit.

\begin{figure}[ht!]
\centering
\includegraphics[scale=1]{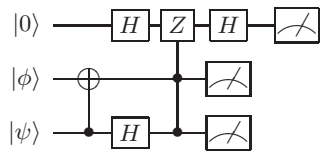}
\caption{SWAP test advancing the measurement.\label{swap4}} 
\end{figure}

To reduce the number of gates, we notice that the ancillary qubit which carries the answer to the test is not affected by the tested qubits after the CCZ gate. We are not interested in the outcomes of the measurements on the qubits under test. We can just as well get rid of the last $H$ and CNOT gates and measure directly after the CCZ gate with no effect on the ancillary qubit and the result of the SWAP test (Figure \ref{swap4}). 
\vspace{2ex}\\

\begin{figure}[ht!]
\centering
\includegraphics[scale=1]{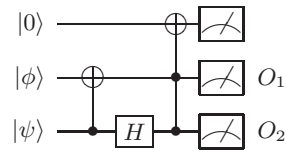}
\caption{SWAP test where the ancillary qubit is the target.\label{swap5}} 
\end{figure}

Now, in the CCZ operation all the qubits can be equally said to be a control or a target. The sign shift only takes place when the three qubits are $\1$. This means we can rewrite the circuit as in Figure \ref{swap5} and make the ancillary qubit the target of a CCNOT gate (using the equivalences in Figure \ref{circuits}).

\begin{figure}[ht!]
\centering
\includegraphics[scale=1]{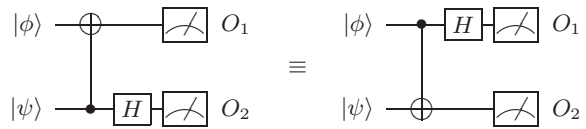}
\caption{Destructive SWAP test.\label{swap6}} 
\end{figure}

The test will fail if, after the CCNOT gate, the original ancillary $\0$ qubit has become $\1$. That only happens when both control qubits are $\1$. We get the same measurement statistics if we just measure the qubits under test and then perform an $X$ gate on the ancillary qubit only if we find two 1 outcomes. This fact is sometimes called the principle of deferred measurement \cite{NC00}. Clearly, we can just ignore the ancillary qubit and perform a SWAP test with the circuit in Figure \ref{swap6}.

The order of the input states is not relevant. The SWAP test should give the same results for $\ph\ps$ and for $\ps\ph$, giving us two equivalent circuits. The result of the SWAP test is the NAND function of the outcomes, $\mbox{NAND}(O_1,O_2)$. Only if both outcomes are 1, $O_1\cdot O_2=1$, we get a failure.

For the rest of the paper, we work with the last circuit in Figure \ref{swap6}. This circuit is, in fact, a measurement in the Bell basis. The gates take inputs from the Bell basis 
\begin{equation}
\{\frac{\ket{00}+\ket{11}}{\sqrt{2}},\frac{\ket{01}+\ket{10}}{\sqrt{2}},\frac{\ket{00}-\ket{11}}{\sqrt{2}},\frac{\ket{01}-\ket{10}}{\sqrt{2}}\}
\end{equation}
into the computational basis  $\{\ket{00},\ket{01},\ket{10},\ket{11}\}$. This is quite relevant, as both the SWAP test and the Hong-Ou-Mandel effect have a peculiar behaviour when the inputs are entangled. Take state $\frac{\ket{01}+\ket{10}}{\sqrt{2}}$. After the change of basis, it becomes $\ket{10}$ and passes the test. This is correct because the entangled input state is the right level of description, but runs against our intuition that it should fail because the first and the second qubit are always different. State comparison is only valid for an input $\ph\ps=\ph\otimes \ps$, where $\otimes$ is a tensor product.

We can do a quick check to find the test is still valid after all the simplifications. For two arbitrary single qubit input states $\ph=\alpha\0+\beta\1$ and $\ps=\gamma\0+\delta\1$, the input state goes from
\begin{equation}
\alpha\gamma\ket{00}+\alpha\delta\ket{01}+\beta\gamma\ket{10}+\beta\delta\ket{11}
\end{equation}
to
\begin{equation}
\alpha\gamma\ket{00}+\alpha\delta\ket{01}+\beta\gamma\ket{11}+\beta\delta\ket{10}
\end{equation}
after the CNOT. After the $H$ gate and before measurement we have
\begin{eqnarray}
\frac{1}{\sqrt{2}}[\alpha\gamma\ket{00}+\alpha\gamma\ket{10}+\alpha\delta\ket{01}+\alpha\delta\ket{11}\nonumber\\
\beta\gamma\ket{01}-\beta\gamma\ket{11}+\beta\delta\ket{00}-\beta\delta\ket{10}].
\end{eqnarray}
The probability of failure $\frac{|\alpha\delta-\beta\gamma|^2}{2}=\frac{(\alpha\delta-\beta\gamma)(\alpha\delta-\beta\gamma)^*}{2}$ comes from considering the probability of measuring the $\ket{11}$ state. The complementary probability of success is
\begin{equation}
P=\frac{2-|\alpha|^2|\delta|^2-|\beta|^2|\gamma|^2+\alpha\beta^*\gamma^*\delta+\alpha^*\beta\gamma\delta^*}{2}.
\end{equation}
The result taking into account the probability amplitudes in the input qubits obey $|\alpha|^2+|\beta|^2=1$ and $|\gamma|^2+|\delta|^2=1$ to obtain
\begin{equation}
P=\frac{1+|\alpha|^2|\gamma|^2+|\beta|^2|\delta|^2+\alpha\beta^*\gamma^*\delta+\alpha^*\beta\gamma\delta^*}{2}.
\label{prob1qubit}
\end{equation}

The overlap of the input states is $|\overl{\psi}{\phi}|^2=(\alpha\gamma^*+\beta\delta^*)(\alpha^*\gamma+\beta^*\delta)=|\alpha|^2|\gamma|^2+|\beta|^2|\delta|^2+\alpha\beta^*\gamma^*\delta+\alpha^*\beta\gamma\delta^*$. We can se the probability of success of the SWAP test $P=\frac{1+|\overl{\psi}{\phi}|^2}{2}$ corresponds to that in Equation (\ref{prob1qubit}).

\subsection{Generalization to $n$ qubits}
The destructive SWAP test can be extended to any number of qubits with little additional effort. We take two $n$-qubit states $\ph$ and $\ps$ so that $\ph\ps=\ph\otimes\ps$. The qubits that form each input state can be entangled.  

\begin{figure}[ht!]
\centering
\includegraphics[scale=1]{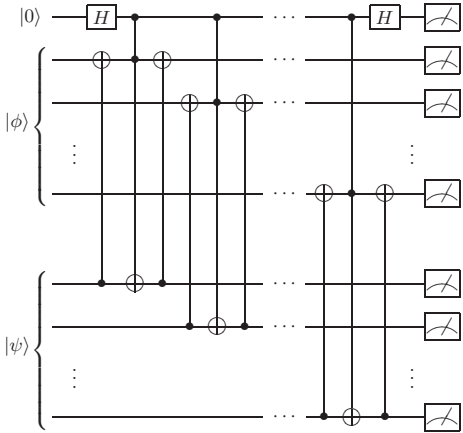}
\caption{SWAP test for $n$-qubit states.\label{nswap}} 
\end{figure}

Clearly, if we swap the qubits of $\ph$ and $\ps$ one by one, we have an $n$-qubit SWAP gate. Figure \ref{nswap} shows the corresponding SWAP test circuit where all the qubits are explicitly shown. 

\begin{figure}[ht!]
\centering
\includegraphics[scale=1]{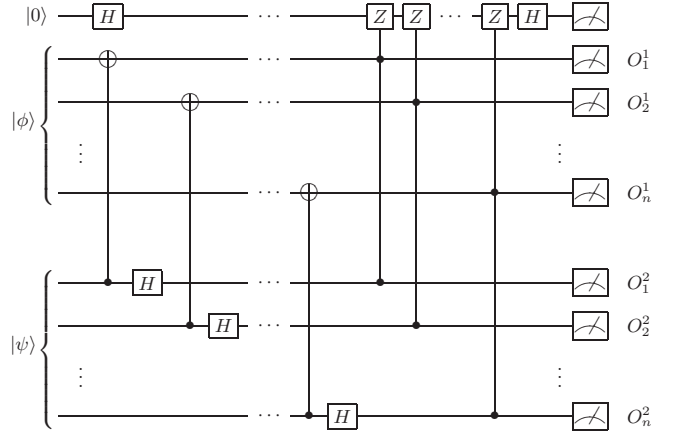}
\caption{SWAP test for $n$-qubit states advancing the measurement.\label{nswap2}} 
\end{figure}

We can repeat the steps of the one qubit states example and get the circuit of Figure \ref{nswap2}. The ancillary qubit sees $n$ CCZ gates. The total phase shift can be perfectly determined from the outcomes of the measurements $O_1^1,\ldots,O_n^1,O_1^2,\ldots,O_n^2$. $O_i^1$ is the result of the measurement on the $i$-th qubit of the first tested state. $O_i^2$ is the corresponding result for the second state. The total phase shift is $\pi\sum_{i=1}^n O_i^1\cdot O_i^1$. The qubit output is 1 (failed test) only if we have an odd number of sign shifts. 

\begin{figure}[ht!]
\centering
\includegraphics[scale=1]{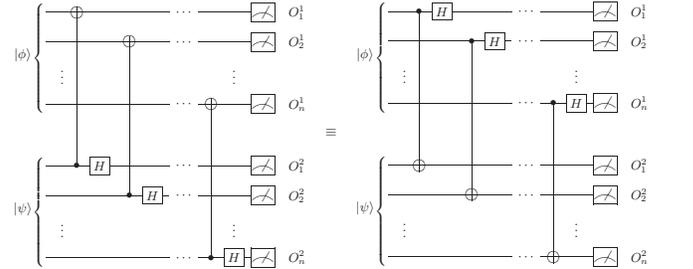}
\caption{SWAP test for $n$-qubit states. We can, once again, change the input state order to obtain the last equivalence.\label{nswap3}} 
\end{figure}

We can ignore the ancillary qubit altogether and obtain the same answer from the measurement outcomes (Figure \ref{nswap3}). If we call $O_1$ and $O_2$ to the bit strings with all the measurements corresponding to all the $O_i^1$ and $O_i^2$, the test succeeds if the bitwise AND of $O_1$ and $O_2$ has an even parity. 

We wish to notice that, while for quantum systems with a dimension that is a power of two there is a natural destructive circuit, the decomposition of the SWAP test circuit for general $d$-dimensional states $\ph$ and $\ps$ (qudits) poses certain challenges. Complement to $d$ and modulo $d$ are not the same operation as in the $d=2$ case. 

\section{An optical SWAP circuit}
\label{opSWAP}
We can also check that the optical circuit of the Hong-Ou-Mandel effect not only performs the same operation as the SWAP test, but is also completely equivalent to the destructive SWAP test of section \ref{dSWAP}.

\subsection{Optical SWAP test}
\begin{figure}[ht!]
\centering
\includegraphics[scale=1]{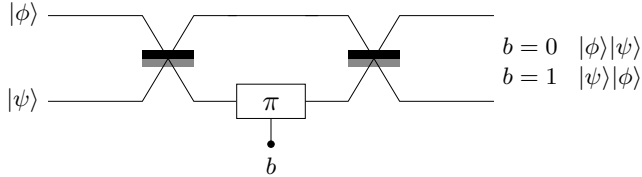}
\caption{Optical SWAP gate with a classical control bit, $b$.\label{ccoptSWAP}} 
\end{figure}

The optical setup of the HOM effect is just a destructive version of the complete optical implementation of the SWAP test. To prove it, we start with the controlled optical SWAP gate in Figure \ref{ccoptSWAP}. The system is a modified interferometer. 

The gate has two 50\% beam splitters and a $\pi$ phase shifter. We add a control bit $b$ that activates the phase shifter when its value is 1. Physically, it can correspong to a Pockels cell, a typical element to manipulate single photons in optical quantum computation \cite{PJF02a}. Pockels cells introduce a $\pi$ phase shift between the upper and lower arms of the interferometer. 

When $b=0$, we have two beam splitters which cancel each other (they apply two H operations in a row). For independent, orthogonal photons, it is clear that, at the second beam splitter, there is a constructive interference in the up or down port the photon came in and a destructive interference in the other port. Taking equations (\ref{BSU}-\ref{BSD}), we can also see that indistinguishable photons are separated again after the second beam splitter. The total evolution is
\begin{equation}
\label{interferom}
\ket{s_U^1}\ket{s_D^1} \stackrel{BS_1}{\longrightarrow} \frac{\ket{s_U^2}\ket{0_D}-\ket{0_U}\ket{s_D^2}}{\sqrt{2}} \stackrel{BS_2}{\longrightarrow} \ket{s_U^1}\ket{s_D^1}.
\end{equation}
Both equal and different components have the same behaviour. We can establish the two beamsplitters perform an identity operation.

When there is a $\pi$ phase shift ($b=1$), we have a typical interferometric setup where the port with the constructive and destructive interference change. For orthogonal photons, we can see from each individual photon's evolution that
\begin{multline}
\ket{s_U^1}\ket{0_D} \stackrel{BS_1}{\longrightarrow} \frac{\ket{s_U^1}\ket{0_D}+\ket{0_U}\ket{s_D^1}}{\sqrt{2}} \\
\stackrel{\pi}{\longrightarrow} \frac{\ket{s_U^1}\ket{0_D}-\ket{0_U}\ket{s_D^1}}{\sqrt{2}} \stackrel{BS_2}{\longrightarrow} \ket{0_U}\ket{s_D^1}
\end{multline}
and
\begin{multline}
\ket{0_U}\ket{s_D^1} \stackrel{BS_1}{\longrightarrow} \frac{\ket{s_U^1}\ket{0_D}-\ket{0_U}\ket{s_D^1}}{\sqrt{2}} \\
\stackrel{\pi}{\longrightarrow} \frac{\ket{s_U^1}\ket{0_D}+\ket{0_U}\ket{s_D^1}}{\sqrt{2}} \stackrel{BS_2}{\longrightarrow} \ket{s_U^1}\ket{0_D}.
\end{multline}
For photons in the same state, we always have 0 or 2 photons going through the phase shifter. This makes a total phase shift of 0 or $2\pi$ for the joint system, which doesn't alter the global state. Equation (\ref{interferom}) is still valid. Anyway, for indistinguishable photons, the output can be equally said to be the same or swapped. 

\begin{figure}[ht!]
\centering
\includegraphics[scale=1]{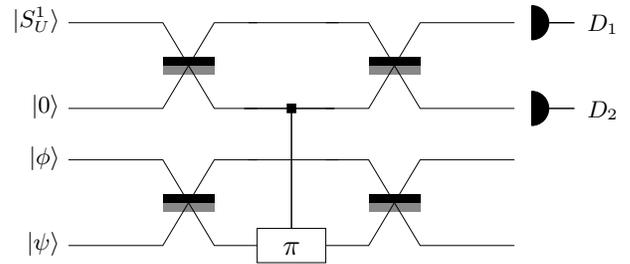}
\caption{Optical SWAP test with an interferometric setup and an optical CZ gate.\label{optSWAP}} 
\end{figure}

We can now add an ancillary photon to perform a full SWAP test (Figure \ref{optSWAP}). This setup is an optical implementation of the circuit in Figure \ref{SWAPcirc}.

We put a photon in any state we want in the upper port of an interferometer with two 50\% beam splitters in the place of the $H$ gates. The most complicated part is the control of the SWAP gate. The logical $\ket{1}$ state, the $\ket{0_U}\ket{s_D^1}$ term after the first beam splitter, must activate the $\pi$ phase shift that triggers the SWAP operation. This is a CZ operation for photons, which, given we can build H gates with beam splitters, is also a photonic CNOT gate. There have been many proposals in that direction, like using the nonlinearities inside Kerr media, or making the photons interact with atomic systems or introducing measurement assisted system \cite{KMN07}. Photonic interaction at the quantum level is challenging and it still remains a major roadblock for scalable optical quantum computation. However, we only need the gate as an intermediate step in our proof. We will assume it is possible to build one and will not really worry about its efficiency.

\begin{figure}[ht!]
\centering
\includegraphics[scale=1]{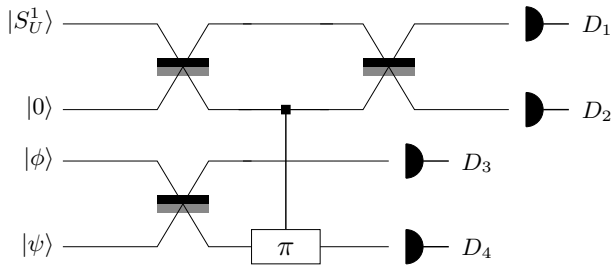}
\caption{Simplified optical SWAP test.\label{optSWAP2}} 
\end{figure}

Now we have all the elements in place, we can proceed in the same way we did in Section \ref{dSWAP}. The input photons pass the SWAP test if detector $D_1$ in Figure \ref{optSWAP} clicks. The two-photon state at the output of the lower interferometer is not used. We could just as well take out the last beam splitter and the SWAP test would be unaffected. We can also add two detectors $D_3$ and $D_4$ (Figure \ref{optSWAP2}).

As we commented in Section \ref{dSWAP}, in a CZ gate the roles of the control and the target states can be reversed. We can suppose the optical CZ gate is controlled by the existence of photons in the lower arm of the lower interferometer. The input photons under comparison fail the test only if there is a $\pi$ phase shift in the lower path of the ancillary photon. Imagine $D_4$ could count photons. For 0 or 2 photons there has been no change in the ancillary photon's phase and we know the SWAP test has been successful. For one photon the input states fail the test. The output state up in the ancillary interferometer is then correlated to the measurement outcomes from detectors $D_3$ and $D_4$.

\begin{figure}[ht!]
\centering
\includegraphics[scale=1]{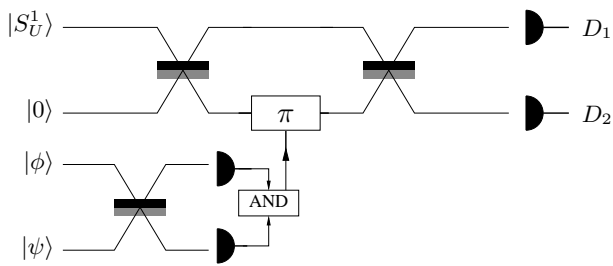}
\caption{Optical SWAP test without the CZ gate.\label{optSWAP3}} 
\end{figure}

The optical CZ gate does not change the number of photons. We can perform the measurement before the gate and get the same measurement statistics (Figure \ref{optSWAP3}). We don't really need to be able to count photons. The total photon number is conserved in the passive, lossless, linear optics beam splitter we are assuming. For two input photons, we have two output photons. The only way to have one photon in $D_4$ is if we get a coincidence count. If only $D_3$ fires, we have two photons up. If only $D_4$ fires, both photons are down. The AND of the outcomes of both detectors, being 0 no click and 1 a click, gives the control bit for the SWAP gate in the upper interferometer. The output of the SWAP test is the NAND of the outcomes. The test fails (outcome 0) only if there is a coincidence count.

\begin{figure}[ht!]
\centering
\includegraphics[scale=1]{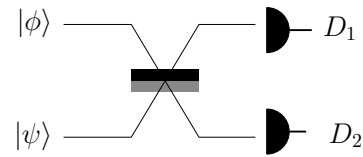}
\caption{Destructive SWAP test with a Hong-Ou-Mandel configuration.\label{optSWAP4}} 
\end{figure}

That means we can just ignore the ancillary photon and work directly with the detectors' outcomes. The usual Hong-Ou-Mandel setup (Figure \ref{optSWAP4}) with simple binary photodetectors that click or not, such as avalanche photodiodes, is enough to perform a SWAP test. All the steps in the proof are valid regardless of the dimension of the photon states $\ph$ and $\ps$. A beam splitter and two photodetectors is all we need to perform a SWAP test on any two photon states.

\section{Applications and future lines}
\label{apps}
We have shown the Hong-Ou-Mandel effect and the SWAP test are formally equivalent. The proof offers simpler implementations of the SWAP test which can be interesting in quantum information protocols. 

Equation (\ref{contHOM}) captures how photons can be used in a SWAP test in quantum information. We only need to find orthogonal wavefunctions. The most obvious examples are frequency and time-bin qudits. The wavefunctions of photons of different frequencies can be thought of as orthogonal sine functions. Time-bin qudits are just wavefunctions with separate, non-overlapping supports. There are also wavefunctions which are orthogonal in space like, for instance, optical vortices carrying Orbital Angular Momentum.

There is a caveat in this last case. Reflection from the beam splitter performs a left-to-right inversion. If we want to preserve proper interference, the reflection must be compensated. Imagine we have input OAM states $\ket{\ell}$ which carry an orbital angular momentum of $\ell\hbar$. In a 50\% beam splitter the evolution is
\begin{eqnarray}
\ket{\ell_U^1}\ket{0_D}&\longrightarrow& \frac{\ket{-\ell_U^1}\ket{0_D}+\ket{0_U}\ket{\ell_D^1}}{\sqrt{2}}\\
\ket{0_U}\ket{\ell_D^1}&\longrightarrow& \frac{\ket{\ell_U^1}\ket{0_D}-\ket{0_U}\ket{-\ell_D^1}}{\sqrt{2}}.
\end{eqnarray}
Due to the symmetry of the OAM wavefronts, reflection from a mirror results in a change of the sign of the winding number $\ell$. A simple mirror in the lower port can compensate for that. An input $\ket{\ell_U^1}\ket{-\ell_D^1}$ becomes, at the output of the beam splitter, the entangled state
\begin{equation}
\frac{\ket{-\ell_U^2}\ket{0_D}-\ket{0_U}\ket{\ell_D^2}}{\sqrt{2}},
\end{equation}
where the interference in the Hong-Ou-Mandel effect is still present and the photons in the upper output port are in the reflected state. A similar analysis can be made for any spatially modulated photon.

One possible application is quantum fingerprinting. Two users, Alice with a string $x$ and Bob with a string $y$, both $n$ bits longs, want to know whether $x$ and $y$ are equal or different. They could send both strings to a referee to compare them. The cost is communicating $2n$ classical bits. Alternatively, they could send shorter strings, called fingerprints, which are a function of $x$ and $y$ and, with high probability, are only equal when $x=y$. Without any previously shared information, Alice and Bob need fingerprints of the order of $\sqrt{n}$ bits \cite{Amb96}. Quantum fingerprints of size of the order of $\log_2(n)$ qubits are enough for the same task \cite{BCW01}. This exponential reduction in communication complexity is based in the comparison of quantum fingerprint states. For a string $x$, the fingerprint is a superposition of $m=cn$ states of the form
\begin{equation}
\label{fingerprint}
\ket{h_x}=\frac{1}{\sqrt{m}}\sum_{i=1}^m (-1)^{E_i(x)}\ket{i}
\end{equation}
where $E(x)$ is the code word corresponding to $x$ in a binary error correcting code and $E_i(x)$ the $i$-th bit of that code word. $c$ is a constant related to the chosen code. For certain error-correcting codes, like Justesen codes \cite{Jus72}, we can guarantee $\overl{h_x}{h_y}\le \delta$ for a $\delta>0$. Repetitions of the SWAP test allow to detect different strings with high probability. In the limit of large $n$, the total communication complexity is of the order of $\log_2(n)$ qubits.  

We can perform the test with a Hong-Ou-Mandel setup. In fact, the equivalence of the SWAP test and the Hong-Ou-Mandel effect has already been noticed in the single qubit case and has been put to use in a quantum fingerprinting scheme with one qubit fingerprints which still has some advantages with respect to any classical method \cite{HBM05}.

Our circuits show the results can be extended to qudits as long as we can still use a single photon for each fingerprint.  The fingerprint can be encoded in a single photon with a method similar to the single photon fingerprinting scheme of Massar \cite{Mas05}. Imagine we take a photon source with a long coherence time. A photon wavefunction of length $T$ seconds can be divided into $m$ parts of duration $\frac{T}{m}$. We can number the portions from 1 to $m$ and define states $\ket{i}$ corresponding to a photon found in the $i$-th segment. The fingerprint state of Equation (\ref{fingerprint}) can be generated with a phase shifter which selectively introduces a $\pi$ shift in the portions for which $E_i(x)$ is 1. Bob can do the same to produce a state $\ket{h_y}$.

While $m$ is of the order of $n$, the photon state can only convey $\log_2 c+\log_2 n$ bits. We can only determine the segment (one out of $m$ possibilites). The Holevo bound makes it impossible to recover more bits \cite{Hol73}. This kind of test would prove the principle of quantum fingerprinting.

However, there are two details that make the system impractical. First, we could use the $T$ seconds to send $x$ directly with classical light using a phase modulation encoding where 0 corresponds to a null phase shift and 1 to a phase $\pi$. The number of bits is greater, but we avoid dealing with single photons and the total transmission time is still $T$. In a practical system, there is no real advantage in using the quantum scheme. Second, in order to obtain a small probability of error the fingerprints have to be sent multiple times. The length of the strings $n$ has to be quite large to beat the communication efficiency the probabilistic classical methods with communication complexity around $\sqrt{n}$ bits. While asymptotically the quantum system is exponentially better, it will only work for long strings around $10^{10}\approx 2^{33}$ \cite{Mas05}, which poses experimental challenges.

The first problem can be solved using better encodings. Hyperentangled photons are a good example \cite{BLP05}. We could use a combination of polarization, OAM and temporal degrees of freedom. For two polarization states, $M$ OAM states and $B$ time-bins, we have $2MB$ orthogonal states. A complete decoding would be difficult, but it is not needed for a SWAP test. Single photon fingerprints in such an encoding would take only $B$ time segments. If $B$ and $M$ are of the same order, close to $\sqrt{\frac{n}{2}}$, we can compete with the classical scheme in terms of the transmission time. 

In that sense, we advocate for spatial encoding schemes. Let's take a spatial light modulator, SLM, with $N\times N$ transmissive pixels which either do nothing or introduce a $\pi$ phase shift. This SLM can produce up to ${2^{N^2}}$ different wavefronts for a single photon. We can search for a subset of those wavefunctions which have a bounded overlap $\overl{h_x}{h_y}\le\delta$ for any $x$ and $y$. This is a generalization of what is done to produce OAM states with SLMs \cite{SGJ07}. Similarly, $d$ states can be encoded in the transverse spatial profile of a single photon \cite{WLA06}. If a good family of codes is found, it would allow to send single photon fingerprints in a reasonable time. Spatial precision needs not to be so good as in a classical spatial encoding method. We just need a binary equal/not equal measurement from the coincidence count. The quantum fingerprinting system is practical as long as we can make the photons interfere (arrange the times of arrival and correct for the effects of reflection in the wavefront). This kind of system would permit many interesting experiments with the SWAP test, not only as used in quantum fingerprinting, but also in other applications such as entanglement detection \cite{HM10}.

The destructive SWAP test of the Hong-Ou-Mandel effect is a practical alternative in simple quantum communication protocols. The optical Hong-Ou-Mandel setup, together with the destructive SWAP circuits, makes an interesting addition to the quantum information toolkit. Reciprocally, the formal equivalence gives us a computational point of view to analyse and understand interference in quantum optics.

\section*{Acknowledgements}
This work has been funded by projects VA342B11-2 (Junta de Castilla y Le\'on) and TEC2010-21303-C04-04 (MICINN).

\newcommand{\noopsort}[1]{} \newcommand{\printfirst}[2]{#1}
  \newcommand{\singleletter}[1]{#1} \newcommand{\switchargs}[2]{#2#1}


\begin{thebibliography}{23}
\expandafter\ifx\csname natexlab\endcsname\relax\def\natexlab#1{#1}\fi
\expandafter\ifx\csname bibnamefont\endcsname\relax
  \def\bibnamefont#1{#1}\fi
\expandafter\ifx\csname bibfnamefont\endcsname\relax
  \def\bibfnamefont#1{#1}\fi
\expandafter\ifx\csname citenamefont\endcsname\relax
  \def\citenamefont#1{#1}\fi
\expandafter\ifx\csname url\endcsname\relax
  \def\url#1{\texttt{#1}}\fi
\expandafter\ifx\csname urlprefix\endcsname\relax\def\urlprefix{URL }\fi
\providecommand{\bibinfo}[2]{#2}
\providecommand{\eprint}[2][]{\url{#2}}

\bibitem[{\citenamefont{Buhrman et~al.}(2010)\citenamefont{Buhrman, Cleve,
  Massar, and de~Wolf}}]{BCM10}
\bibinfo{author}{\bibfnamefont{H.}~\bibnamefont{Buhrman}},
  \bibinfo{author}{\bibfnamefont{R.}~\bibnamefont{Cleve}},
  \bibinfo{author}{\bibfnamefont{S.}~\bibnamefont{Massar}}, \bibnamefont{and}
  \bibinfo{author}{\bibfnamefont{R.}~\bibnamefont{de~Wolf}},
  \bibinfo{journal}{Rev. Mod. Phys.} \textbf{\bibinfo{volume}{82}},
  \bibinfo{pages}{665} (\bibinfo{year}{2010}).

\bibitem[{\citenamefont{Bennett and Brassard}(1984)}]{BB84}
\bibinfo{author}{\bibfnamefont{C.~H.} \bibnamefont{Bennett}} \bibnamefont{and}
  \bibinfo{author}{\bibfnamefont{G.}~\bibnamefont{Brassard}},
  \bibinfo{journal}{Proceedings of IEEE international Conference on Computers,
  Systems and Signal Processing, Bangalore, India} p. \bibinfo{pages}{175}
  (\bibinfo{year}{1984}).

\bibitem[{\citenamefont{Gisin et~al.}(2002)\citenamefont{Gisin, Ribordy,
  Tittel, and Zbinden}}]{GRT02}
\bibinfo{author}{\bibfnamefont{N.}~\bibnamefont{Gisin}},
  \bibinfo{author}{\bibfnamefont{G.}~\bibnamefont{Ribordy}},
  \bibinfo{author}{\bibfnamefont{W.}~\bibnamefont{Tittel}}, \bibnamefont{and}
  \bibinfo{author}{\bibfnamefont{H.}~\bibnamefont{Zbinden}},
  \bibinfo{journal}{Rev. Mod. Phys.} \textbf{\bibinfo{volume}{74}},
  \bibinfo{pages}{145} (\bibinfo{year}{2002}).

\bibitem[{\citenamefont{Grover}(2001)}]{Gro01}
\bibinfo{author}{\bibfnamefont{L.~K.} \bibnamefont{Grover}},
  \bibinfo{journal}{American Journal of Physics} \textbf{\bibinfo{volume}{69}},
  \bibinfo{pages}{769} (\bibinfo{year}{2001}).

\bibitem[{\citenamefont{Buhrman et~al.}(2001)\citenamefont{Buhrman, Cleve,
  Watrous, and de~Wolf}}]{BCW01}
\bibinfo{author}{\bibfnamefont{H.}~\bibnamefont{Buhrman}},
  \bibinfo{author}{\bibfnamefont{R.}~\bibnamefont{Cleve}},
  \bibinfo{author}{\bibfnamefont{J.}~\bibnamefont{Watrous}}, \bibnamefont{and}
  \bibinfo{author}{\bibfnamefont{R.}~\bibnamefont{de~Wolf}},
  \bibinfo{journal}{Physical Review Letters} \textbf{\bibinfo{volume}{87}},
  \bibinfo{pages}{167902} (\bibinfo{year}{2001}).

\bibitem[{\citenamefont{Holevo}(1973)}]{Hol73}
\bibinfo{author}{\bibfnamefont{A.}~\bibnamefont{Holevo}},
  \bibinfo{journal}{Problems of Information Transmission}
  \textbf{\bibinfo{volume}{9}}, \bibinfo{pages}{177} (\bibinfo{year}{1973}).

\bibitem[{\citenamefont{Holevo}(1998)}]{Hol98}
\bibinfo{author}{\bibfnamefont{A.}~\bibnamefont{Holevo}},
  \bibinfo{journal}{Information Theory, IEEE Transactions on}
  \textbf{\bibinfo{volume}{44}}, \bibinfo{pages}{269} (\bibinfo{year}{1998}).

\bibitem[{\citenamefont{Hong et~al.}(1987)\citenamefont{Hong, Ou, and
  Mandel}}]{HOM87}
\bibinfo{author}{\bibfnamefont{C.~K.} \bibnamefont{Hong}},
  \bibinfo{author}{\bibfnamefont{Z.~Y.} \bibnamefont{Ou}}, \bibnamefont{and}
  \bibinfo{author}{\bibfnamefont{L.}~\bibnamefont{Mandel}},
  \bibinfo{journal}{Phys. Rev. Lett.} \textbf{\bibinfo{volume}{59}},
  \bibinfo{pages}{2044} (\bibinfo{year}{1987}).

\bibitem[{\citenamefont{Loudon}(2000)}]{Lou00}
\bibinfo{author}{\bibfnamefont{R.}~\bibnamefont{Loudon}},
  \emph{\bibinfo{title}{The Quantum Theory of Light}}
  (\bibinfo{publisher}{Oxford University Press}, \bibinfo{address}{Great
  Clarendon Street, Oxford, UK}, \bibinfo{year}{2000}), \bibinfo{edition}{3rd}
  ed.

\bibitem[{\citenamefont{Skaar et~al.}(2004)\citenamefont{Skaar, Escart\'{\i}n,
  and Landro}}]{SGL04}
\bibinfo{author}{\bibfnamefont{J.}~\bibnamefont{Skaar}},
  \bibinfo{author}{\bibfnamefont{J.~C.~G.} \bibnamefont{Escart\'{\i}n}},
  \bibnamefont{and} \bibinfo{author}{\bibfnamefont{H.}~\bibnamefont{Landro}},
  \bibinfo{journal}{American Journal of Physics} \textbf{\bibinfo{volume}{72}},
  \bibinfo{pages}{1385} (\bibinfo{year}{2004}).

\bibitem[{\citenamefont{Molina-Terriza
  et~al.}(2001)\citenamefont{Molina-Terriza, Torres, and Torner}}]{MTT01}
\bibinfo{author}{\bibfnamefont{G.}~\bibnamefont{Molina-Terriza}},
  \bibinfo{author}{\bibfnamefont{J.~P.} \bibnamefont{Torres}},
  \bibnamefont{and} \bibinfo{author}{\bibfnamefont{L.}~\bibnamefont{Torner}},
  \bibinfo{journal}{Physical Review Letters} \textbf{\bibinfo{volume}{88}},
  \bibinfo{pages}{013601} (\bibinfo{year}{2001}).

\bibitem[{\citenamefont{{Stucki} et~al.}(2005)\citenamefont{{Stucki},
  {Zbinden}, and {Gisin}}}]{SZG05}
\bibinfo{author}{\bibfnamefont{D.}~\bibnamefont{{Stucki}}},
  \bibinfo{author}{\bibfnamefont{H.}~\bibnamefont{{Zbinden}}},
  \bibnamefont{and} \bibinfo{author}{\bibfnamefont{N.}~\bibnamefont{{Gisin}}},
  \bibinfo{journal}{Journal of Modern Optics} \textbf{\bibinfo{volume}{52}},
  \bibinfo{pages}{2637} (\bibinfo{year}{2005}).

\bibitem[{\citenamefont{Nielsen and Chuang}(2000)}]{NC00}
\bibinfo{author}{\bibfnamefont{M.~A.} \bibnamefont{Nielsen}} \bibnamefont{and}
  \bibinfo{author}{\bibfnamefont{I.~L.} \bibnamefont{Chuang}},
  \emph{\bibinfo{title}{Quantum Computation and Quantum Information}}
  (\bibinfo{publisher}{Cambridge University Press},
  \bibinfo{address}{Cambridge, UK}, \bibinfo{year}{2000}),
  \bibinfo{edition}{1st} ed.

\bibitem[{\citenamefont{Pittman et~al.}(2002)\citenamefont{Pittman, Jacobs, and
  Franson}}]{PJF02a}
\bibinfo{author}{\bibfnamefont{T.~B.} \bibnamefont{Pittman}},
  \bibinfo{author}{\bibfnamefont{B.~C.} \bibnamefont{Jacobs}},
  \bibnamefont{and} \bibinfo{author}{\bibfnamefont{J.~D.}
  \bibnamefont{Franson}}, \bibinfo{journal}{Physical Review A}
  \textbf{\bibinfo{volume}{66}}, \bibinfo{pages}{052305}
  (\bibinfo{year}{2002}).

\bibitem[{\citenamefont{Kok et~al.}(2007)\citenamefont{Kok, Munro, Nemoto,
  Ralph, Dowling, and Milburn}}]{KMN07}
\bibinfo{author}{\bibfnamefont{P.}~\bibnamefont{Kok}},
  \bibinfo{author}{\bibfnamefont{W.~J.} \bibnamefont{Munro}},
  \bibinfo{author}{\bibfnamefont{K.}~\bibnamefont{Nemoto}},
  \bibinfo{author}{\bibfnamefont{T.~C.} \bibnamefont{Ralph}},
  \bibinfo{author}{\bibfnamefont{J.~P.} \bibnamefont{Dowling}},
  \bibnamefont{and} \bibinfo{author}{\bibfnamefont{G.~J.}
  \bibnamefont{Milburn}}, \bibinfo{journal}{Reviews of Modern Physics}
  \textbf{\bibinfo{volume}{79}}, \bibinfo{eid}{135}
  (pages~\bibinfo{numpages}{40}) (\bibinfo{year}{2007}).

\bibitem[{\citenamefont{Ambainis}(1996)}]{Amb96}
\bibinfo{author}{\bibfnamefont{A.}~\bibnamefont{Ambainis}},
  \bibinfo{journal}{Algorithmica} \textbf{\bibinfo{volume}{16}},
  \bibinfo{pages}{298} (\bibinfo{year}{1996}).

\bibitem[{\citenamefont{Justesen}(1972)}]{Jus72}
\bibinfo{author}{\bibfnamefont{J.}~\bibnamefont{Justesen}},
  \bibinfo{journal}{Information Theory, IEEE Transactions on}
  \textbf{\bibinfo{volume}{18}}, \bibinfo{pages}{652} (\bibinfo{year}{1972}).

\bibitem[{\citenamefont{Horn et~al.}(2005)\citenamefont{Horn, Babichev,
  Marzlin, Lvovsky, and Sanders}}]{HBM05}
\bibinfo{author}{\bibfnamefont{R.~T.} \bibnamefont{Horn}},
  \bibinfo{author}{\bibfnamefont{S.~A.} \bibnamefont{Babichev}},
  \bibinfo{author}{\bibfnamefont{K.-P.} \bibnamefont{Marzlin}},
  \bibinfo{author}{\bibfnamefont{A.~I.} \bibnamefont{Lvovsky}},
  \bibnamefont{and} \bibinfo{author}{\bibfnamefont{B.~C.}
  \bibnamefont{Sanders}}, \bibinfo{journal}{Physical Review Letters}
  \textbf{\bibinfo{volume}{95}}, \bibinfo{pages}{150502}
  (\bibinfo{year}{2005}).

\bibitem[{\citenamefont{Massar}(2005)}]{Mas05}
\bibinfo{author}{\bibfnamefont{S.}~\bibnamefont{Massar}},
  \bibinfo{journal}{Physical Review A} \textbf{\bibinfo{volume}{71}},
  \bibinfo{pages}{012310} (\bibinfo{year}{2005}).

\bibitem[{\citenamefont{Barreiro et~al.}(2005)\citenamefont{Barreiro, Langford,
  Peters, and Kwiat}}]{BLP05}
\bibinfo{author}{\bibfnamefont{J.~T.} \bibnamefont{Barreiro}},
  \bibinfo{author}{\bibfnamefont{N.~K.} \bibnamefont{Langford}},
  \bibinfo{author}{\bibfnamefont{N.~A.} \bibnamefont{Peters}},
  \bibnamefont{and} \bibinfo{author}{\bibfnamefont{P.~G.} \bibnamefont{Kwiat}},
  \bibinfo{journal}{Physical Review Letters} \textbf{\bibinfo{volume}{95}},
  \bibinfo{eid}{260501} (\bibinfo{year}{2005}).

\bibitem[{\citenamefont{St\"{u}tz et~al.}(2007)\citenamefont{St\"{u}tz,
  Gr\"{o}blacher, Jennewein, and Zeilinger}}]{SGJ07}
\bibinfo{author}{\bibfnamefont{M.}~\bibnamefont{St\"{u}tz}},
  \bibinfo{author}{\bibfnamefont{S.}~\bibnamefont{Gr\"{o}blacher}},
  \bibinfo{author}{\bibfnamefont{T.}~\bibnamefont{Jennewein}},
  \bibnamefont{and}
  \bibinfo{author}{\bibfnamefont{A.}~\bibnamefont{Zeilinger}},
  \bibinfo{journal}{Applied Physics Letters} \textbf{\bibinfo{volume}{90}},
  \bibinfo{eid}{261114} (\bibinfo{year}{2007}).

\bibitem[{\citenamefont{Walborn et~al.}(2006)\citenamefont{Walborn, Lemelle,
  Almeida, and Ribeiro}}]{WLA06}
\bibinfo{author}{\bibfnamefont{S.~P.} \bibnamefont{Walborn}},
  \bibinfo{author}{\bibfnamefont{D.~S.} \bibnamefont{Lemelle}},
  \bibinfo{author}{\bibfnamefont{M.~P.} \bibnamefont{Almeida}},
  \bibnamefont{and} \bibinfo{author}{\bibfnamefont{P.~H.}
  \bibnamefont{Souto Ribeiro}}, \bibinfo{journal}{Physical Review Letters}
  \textbf{\bibinfo{volume}{96}}, \bibinfo{eid}{090501}
 (\bibinfo{year}{2006}).

\bibitem[{\citenamefont{Harrow and Montanaro}(2010)}]{HM10}
\bibinfo{author}{\bibfnamefont{A.}~\bibnamefont{Harrow}} \bibnamefont{and}
  \bibinfo{author}{\bibfnamefont{A.}~\bibnamefont{Montanaro}}, in
  \emph{\bibinfo{booktitle}{Foundations of Computer Science (FOCS), 2010 51st
  Annual IEEE Symposium on}} (\bibinfo{year}{2010}), pp.
  \bibinfo{pages}{633--642}.

\end{thebibliography}
\end{document}